\documentclass[sigconf,nonacm]{acmart}
\renewcommand\footnotetextcopyrightpermission[1]{}
\AtBeginDocument{%
  }

\begin{document}

\title{Q-ARE: An Evaluation Dataset for Query Based API Recommendation}

\author{Shenglong Wu}
\affiliation{%
  \institution{College of Computer Science and Technology, National University of Defense Technology}
  \city{Changsha}
  \state{Hunan}
  \country{China}
}
\affiliation{%
  \institution{State Key Laboratory of Complex \& Critical Software Environment}
  \city{Changsha}
  \state{Hunan}
  \country{China}
}
\email{wushenglong23@nudt.edu.cn}

\author{Xunhui Zhang}

\affiliation{%
  \institution{College of Computer Science and Technology, National University of Defense Technology}
  \city{Changsha}
  \state{Hunan}
  \country{China}
}
\affiliation{%
  \institution{State Key Laboratory of Complex \& Critical Software Environment}
  \city{Changsha}
  \state{Hunan}
  \country{China}
}
\email{zhangxunhui@nudt.edu.cn}

\author{Tao Wang}
\authornote{This is the corresponding author.}
\affiliation{%
  \institution{College of Computer Science and Technology, National University of Defense Technology}
  \city{Changsha}
  \state{Hunan}
  \country{China}
}
\affiliation{%
  \institution{State Key Laboratory of Complex \& Critical Software Environment}
  \city{Changsha}
  \state{Hunan}
  \country{China}
}
\email{taowang2005@outlook.com}

\renewcommand{\shortauthors}{Wu et al.}

\begin{abstract}
\quad As software systems grow in scale, developers face increasing difficulty in selecting appropriate Application Programming Interfaces (APIs) from numerous options. Efficiently identifying APIs that satisfy functional requirements has become a key challenge. To evaluate the semantic understanding of existing query-based API recommendation methods, this paper constructs \textbf{Q-ARE (Query-based API Recommendation Evaluation)}, a dataset based on open-source Java projects from GitHub. Methods and their invocation chains are analyzed to identify third-party APIs directly or indirectly invoked by target methods, recursively expanding \textbf{multi-level} invocations to unify hierarchical call structures into API recommendation target sets. Furthermore, we introduce two metrics: \textbf{API Call Depth}, measuring the invocation distance between a query method and a target API, and Invocation Density, quantifying the proportion of code lines associated with the target API in the invocation chain. Based on \textbf{Q-ARE}, we systematically evaluate several query-based API recommendation methods and general \textbf{Large Language Models (LLMs)}. Results show that performance drops significantly as \textbf{API Call Depth} increases and invocation density decreases, indicating that existing methods still struggle with \textbf{multi-level method invocation structures}. \textbf{Q-ARE} and its metrics provide a new benchmark for assessing semantic understanding in API recommendation and offer insights for improving future algorithms.
\end{abstract}



\maketitle

\section{Introduction}

Application Programming Interfaces (APIs) are a fundamental component of modern software development. By invoking APIs, developers can reuse existing software functionalities without implementing complex system modules from scratch, thereby improving development efficiency and software quality~\cite{wang2020empirical}. For example, many third-party libraries and frameworks expose functionalities such as data processing, network communication, and database access through APIs. With the rapid growth of the open-source ecosystem, a large number of feature-rich APIs have been continuously released and widely adopted in software projects. Proper use of these APIs can significantly improve development efficiency, reduce development costs, and enhance the reliability and maintainability of software systems.

However, as the software ecosystem continues to expand, the number of available APIs has grown rapidly, making it increasingly
difficult for developers to quickly identify suitable APIs that satisfy their requirements during development~\cite{peng2022revisiting}. To address this challenge, researchers have proposed API recommendation techniques, which aim to automatically recommend relevant APIs based on developers’ natural language queries or code contexts~\cite{huang2018api,rahman2016rack,wei2022clear,chen2021holistic,ling2019graph}, thereby assisting developers in completing programming tasks more efficiently. In recent years, various API recommendation approaches have been proposed, including information retrieval--based methods~\cite{irsan2023picaso}, deep learning--based methods~\cite{feng2020codebert,gu2016deep}, and approaches leveraging Large Language Models (LLMs)~\cite{kang2021apirecx}. These methods have improved the effectiveness of API recommendation to some extent.

\begin{figure}[ht]
\centering
\includegraphics[width=0.7\linewidth]{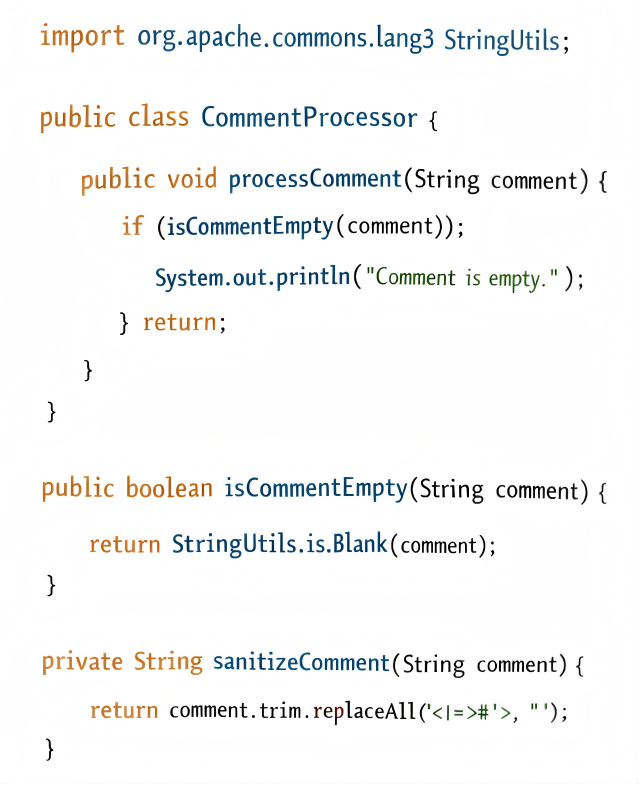} 
\caption{Example of multi-level invocation of a third-party API}
\label{fig:indirect_api_call}
\vspace{-1.1em} 
\end{figure}

Despite the progress made in API recommendation research, the evaluation of existing approaches typically relies on datasets constructed from code snippets, API documentation, or question-answering communities~\cite{peng2022revisiting,liu2024toolace}. 
These datasets mainly focus on APIs that are invoked in a single-level manner. However, in real-world software projects, third-party APIs are often invoked in a multi-level manner through method invocation chains. As shown in Figure~\ref{fig:indirect_api_call}, the method \texttt{processComment} does not invoke the \texttt{isBlank} API from the Apache Commons Lang library in a single-level manner; instead, it calls the API in a multi-level manner through the \texttt{isCommentEmpty} method.

To address this limitation, we mine real-world open-source Java projects from GitHub and construct a new query-based API recommendation evaluation dataset, named \textbf{Q-ARE}. We first select projects satisfying quality criteria. Then, by analyzing both the invocation relationships among methods and between methods and APIs, we recursively resolve method invocation chains to identify third-party APIs that are directly or indirectly invoked by the target methods.Furthermore, we expand multi-level method invocation relationships and transform hierarchical call structures into a unified representation consisting of query methods and the set of reachable third-party APIs, constructing a unified API recommendation target space.

Based on this dataset, we introduce two metrics: \textbf{API Call Depth} and \textbf{API Call Density}. \textbf{Depth}  characterizes the invocation distance between a query method and a target API. Density measures the semantic proportion of target APIs within the code associated with the invocation chain, quantified by the code size of the methods involved. When the call chain involves more code, the semantic contribution of the API within the overall method logic is smaller, increasing the difficulty of the API recommendation task.

Using the proposed dataset, we conduct a systematic evaluation of several query-based API recommendation approaches and general LLMs. Experimental results show that recommendation performance decreases significantly as \textbf{API Call Depth} increases and semantic density decreases. These findings indicate that existing methods still struggle to capture semantic relationships across method invocations and complex functional query, especially when APIs are used indirectly through multi-level invocation chains.

The main contributions of this work are as follows:

1. We construct a query-based API recommendation evaluation dataset, \textbf{Q-ARE}, and provide its full source code for public use. This dataset is built from real open-source projects on GitHub and includes both direct and indirect third-party API invocations through method invocation analysis. It also provides corresponding semantic information, thereby reflecting the actual usage of APIs in software projects more realistically. The dataset and source code are available at \href{https://doi.org/10.5281/zenodo.19907209}{https://doi.org/10.5281/zenodo.19907209}.

2. We propose two metrics: \textbf{API Call Depth} and \textbf{API Call Density}. \textbf{Depth} quantifies the invocation distance between a query method and a target API, while \textbf{Density} measures the sparsity of the target API in the method implementation, i.e., the proportion of code in the invocation chain occupied by the API, reflecting its significance in the overall implementation logic.

3. We systematically evaluate existing API recommendation methods and large language models on the \textbf{Q-ARE} dataset. Experimental results reveal that \textbf{API Call Depth} and semantic proportion have a significant impact on recommendation performance and indicate that current methods still exhibit substantial limitations in complex semantic understanding and cross-method reasoning.

The rest of the paper is organized as follows. Chapter 2 reviews related work. Chapter 3 describes the dataset construction. Chapter 4 presents the experimental setup and results. Chapter 5 discusses the findings and future directions. Chapter 6 examines threats to validity. Finally, Chapter 7 concludes the paper.

\section{Related Work}

\subsection{Query-Based API Recommendation}

Query-based API recommendation aims to automatically suggest APIs that can achieve a target functionality based on natural language descriptions provided by developers. According to APIGen~\cite{chen2024apigen}, existing approaches can be broadly categorized into retrieval-based methods, learning-based methods, and LLM-based approaches.

\paragraph{Retrieval-based methods.} Retrieval-based methods typically recommend APIs by computing the semantic similarity between a query and API documentation or code descriptions. These methods usually first construct semantic representations of API documentation or code snippets and then retrieve candidate APIs based on similarity. For example, CLEAR~\cite{wei2022clear} employs RoBERTa to learn semantic representations of natural language queries and API descriptions, and further optimizes the embedding space using contrastive learning. After retrieving candidate APIs, a BERT-based binary classification model is applied for re-ranking, achieving high recommendation accuracy. Such methods are relatively simple to implement and highly interpretable, but their performance generally depends on the quality of API documentation and the semantic alignment between queries and documentation, limiting their ability to handle complex semantics or indirect API usage scenarios.

\paragraph{Learning-based methods.} Learning-based methods typically model API recommendation as a sequence generation or ranking task, leveraging deep learning models to learn the mapping between queries and APIs from large-scale historical code or Q\&A data. For instance, DeepAPI~\cite{gu2016deep} adopts an RNN encoder--decoder architecture to encode natural language queries into semantic vectors and generates the corresponding API call sequences via the decoder. The method incorporates an attention mechanism, enabling the model to dynamically focus on different parts of the input query during generation, thereby improving the accuracy of API sequence generation. Moreover, BRAID~\cite{zhou2021boosting} proposes a plug-and-play enhancement framework that learns developers’ implicit feedback on recommended results and combines it with active learning to re-rank outputs from existing recommendation systems, significantly improving Top-1 recommendation performance. Compared to retrieval-based methods, learning-based approaches can capture more complex semantic relationships, but their effectiveness often depends on the size of the training data and may be constrained by limitations in modeling long-range dependencies.

\paragraph{LLM-based approaches.} With the rapid advancement of large language models (LLMs) in code understanding and generation, researchers have explored LLM-based API recommendation. For example, APIGen~\cite{chen2024apigen} retrieves historical programming Q\&A semantically similar to a user query as examples and combines them with highly relevant “API seeds” extracted from official documentation to guide LLMs in generating semantically relevant API call sequences, achieving generative API recommendation. Compared with traditional methods, LLMs demonstrate stronger potential in semantic understanding and code generation, yet their effectiveness in recommending APIs for complex software projects still requires systematic evaluation.

In summary, although existing query-based API recommendation methods model the semantic relationships between queries and third-party APIs from various perspectives, constructing an evaluation dataset that can systematically assess the semantic understanding capabilities of these methods is of significant research value and can provide a benchmark for improving future approaches.

\subsection{API Recommendation Evaluation Datasets}

The development of API recommendation methods largely relies on high-quality evaluation datasets. Existing datasets are primarily derived from open-source software projects ~\cite{gu2018deep,gu2016deep,wang2024systematic}, Q\&A communities ~\cite{rahman2016rack,huang2018api,peng2022revisiting}, and manually constructed data ~\cite{song2025callnavi,chen2024apigen,liu2025think}.

A dataset from Gu et al.~\cite{gu2016deep} is built from Java projects on GitHub. Methods with non-empty Javadoc comments ending with a period were selected, and the first sentence of each Javadoc was used as a natural language description. Static analysis of method bodies extracted all external API calls (e.g., \texttt{java.io.File.<init>}, \texttt{DocumentBuilder.parse}) to form API sequences. After cleaning, the dataset contains around 7.5 million method–API pairs, covering Java standard libraries and third-party APIs.

\textbf{APIBENCH-Q}~\cite{peng2022revisiting} evaluates query-based API recommendation methods using data from Stack Overflow and tutorial websites (e.g., GeeksforGeeks, Java2s, Kode Java). Stack Overflow posts were filtered by rules (removing posts without answers, code snippets, or relevant API calls) and manually verified in two rounds to ensure queries expressed clear functional requirements and answers contained corresponding API calls. Tutorial website data were filtered using rules only. The dataset contains 6,563 Java queries and 4,309 Python queries, each paired with accurate ground-truth APIs.

\textbf{THINK}~\cite{liu2025think} collects 150 real-world programming tasks from GitHub and CodeGen4Libs, supplemented with 60 GPT-4-generated tasks, totaling 210 tasks. These cover Java third-party libraries in PDF, image, and NLP domains, involving 3,234 classes and 25,451 APIs, providing diverse and realistic evaluation scenarios.

Existing datasets have limitations. Most focus on single-level API calls, whereas real projects often involve multi-level invocation chains. Some consider only tasks completed by a single API, not capturing multi-API cooperation. While some datasets account for parameter dependencies, their data are mainly LLM-generated or manually constructed and do not fully reflect complex real-world invocation structures.

To address these issues, we construct the \textbf{Q-ARE} dataset from real-world GitHub Java projects. By analyzing method invocations and recursively resolving chains, it identifies both single- and multi-level API calls. We introduce two evaluation metrics: \textbf{API Call Depth}, quantifying the distance between a query method and a target API, and \textbf{API Call Density}, measuring the proportion of code lines associated with the target API in the invocation chain. These metrics provide a systematic benchmark for evaluating API recommendation methods and LLMs in complex invocation scenarios.

\section{Dataset Construction}

\begin{figure*}[t]
  \centering
  \includegraphics[width=0.9\textwidth]{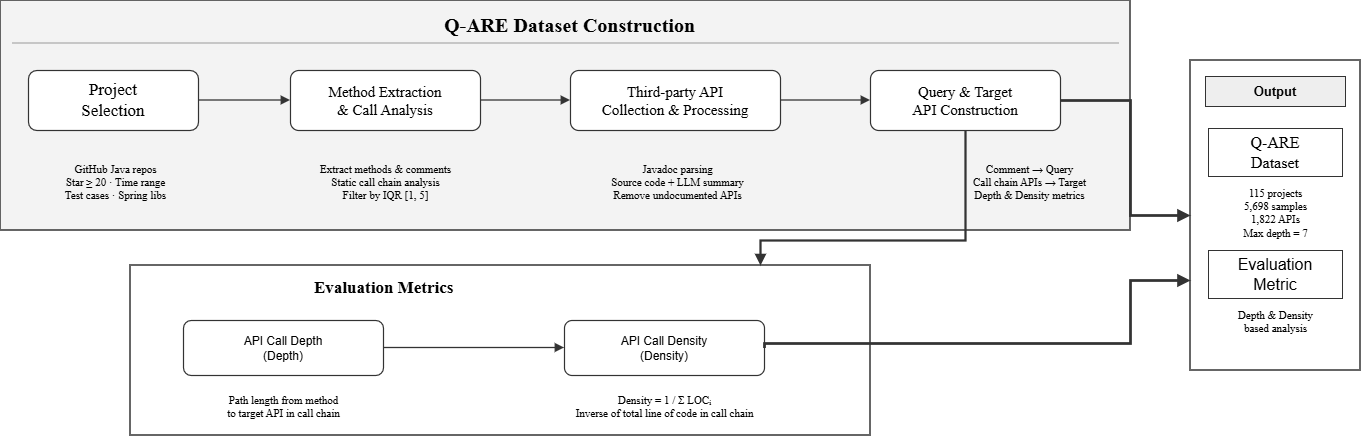}
  \caption{Dataset Construction Pipeline}
  \label{fig:qare_pipeline}
\end{figure*}

To systematically evaluate the performance of query-based API recommendation methods and general-purpose \textbf{LLMs} under complex call structures and semantic contexts, we construct a new query-based API recommendation evaluation dataset, \textbf{Q-ARE}. The dataset construction process consists of four main stages: \textbf{Project Selection}, \textbf{Method Extraction and Call Relation Analysis}, \textbf{Third-Party API Collection and Processing}, and \textbf{Query–Target API Construction}, as illustrated in Figure~\ref{fig:qare_pipeline}.

\subsection{Project Selection}

To ensure the representativeness, quality, and executability of the projects in the dataset, we selected GitHub open-source Java projects that meet the following criteria:

1. \textbf{GitHub Stars ≥ 20}: The number of stars reflects a project's community attention and activity, and to some extent, its maintenance quality. Based on this consideration, and drawing on the authors’ extensive experience in software development and open-source projects, we reached a consensus to use a GitHub Star threshold of 20 or more as an initial screening criterion, ensuring that the selected projects have a certain level of community recognition and practical value.

2. \textbf{Creation Date between February 6, 2025 and December 31, 2025}: The starting date is determined with reference to the release timeline of mainstream large language models. Specifically, we first construct a baseline set using representative models from the OpenCompass LLM Leaderboard~\cite{OpenCompass2024}, and take the most recently released model among them (Gemini-2.0-Pro-Exp-02-05, released on February 5, 2025) as the lower bound of the timeline. Accordingly, the following day (i.e., February 6, 2025) is used as the starting point for data filtering. The end date is chosen to ensure that all selected projects were created before the start of the experiments (January 1, 2026) and have accumulated sufficient development and evolution history to support subsequent analysis and evaluation.

3. \textbf{Contains Test Cases}: Select projects that include executable test cases to support standardized verification of code functionality and to provide a foundation for automated evaluation in subsequent tasks (e.g., method generation based on API recommendation results).

4. \textbf{Use of Mainstream Third-Party Libraries (e.g., Spring Framework)}: This criterion ensures that projects contain rich third-party API usage, providing sufficient samples for evaluating API recommendation methods. Spring Framework is one of the most widely used frameworks in web development and research; its comprehensive documentation guarantees clear API descriptions, which facilitates accurate and in-depth evaluation and ensures that experimental results are representative of realistic scenarios.

Through these selection criteria, we obtained a set of representative real-world software projects, providing a reliable data source for constructing the \textbf{Q-ARE} dataset.

\subsection{Method Extraction and Invocation Relationship Analysis}
Within the selected projects, we perform static analysis based on Eclipse JDT (Java Development Tools) to construct abstract syntax trees (ASTs) from the source code. We first extract all methods along with their corresponding comment information. Considering that natural language queries typically originate from method descriptions, only methods containing comment information are retained as candidate samples. For comments written in Chinese, large language models are used to translate them into English to ensure consistency in semantic representation.

Next, we perform static analysis on the project code to resolve invocation relationships both among methods and between methods and APIs. We first construct a set of third-party libraries by aggregating all imported third-party libraries across the projects, and use this set to identify third-party APIs involved in method calls. Calls that do not belong to this set (e.g., internal project methods or standard library calls) are filtered out.

Furthermore, we count the number of third-party APIs called by each method and use boxplots to visualize their distribution. The results show that the interquartile range (IQR) of API calls is [2, 5]. To focus on methods with typical external dependency structures while avoiding noise caused by extremely sparse or highly coupled methods, only methods with API call counts within this IQR are retained for analysis.

Based on this, we recursively resolve method invocation chains to identify third-party APIs directly or indirectly invoked by the target methods, providing the foundational data for constructing the mapping between queries and target APIs.

\subsection{Third-Party API Collection and Processing}

After parsing call chains, all directly or indirectly invoked third-party APIs were collected and deduplicated to form the target API set. To support semantic matching in API recommendation tasks, textual descriptions of APIs were also collected using the following steps:

\begin{itemize}
    \item \textbf{Javadoc parsing:} Extract official documentation comments from the library's \texttt{javadoc.jar}.
    \item \textbf{Source code parsing and summary generation:} For APIs lacking documentation, source code was obtained from \texttt{sources.jar} or by decompiling JAR files, and concise functional descriptions were generated using LLMs.
    \item \textbf{Filtering strategy:} APIs with neither documentation nor parseable source code were removed, and any methods dependent on these APIs were also discarded.
\end{itemize}

This process ensures that the \textbf{Q-ARE} dataset captures method dependencies on third-party APIs, including both direct and multi-level indirect calls, in real software projects.

\subsection{Query and Target API Construction}

During the dataset construction phase, we treat method comments as natural language queries (\textbf{Query}) and the third-party API sets involved in method invocation chains as target API sets (\textbf{Target APIs}), thereby establishing a mapping between queries and APIs. To evaluate the complexity of API usage structures and the deep semantic understanding capabilities of baseline methods, we introduce two metrics:

\begin{itemize}
  \item \textbf{API Call Depth} measures the invocation distance between a method and its target API. Specifically, it represents the path length from the method to the target API along the invocation chain. For example, the method \texttt{process\allowbreak Comment} calls \texttt{isCommentEmpty}, which in turn invokes the third-party API \texttt{isBlank}. In this case, the call depth of \texttt{isBlank} relative to \texttt{process\allowbreak Comment} is 3, while its depth relative to \texttt{isCommentEmpty} is 2.

  \item \textbf{API Call Density} quantifies the proportion of lines of code associated with a target API within the relevant invocation chain. Specifically, we compute the total lines of code (LOC) across all methods in the invocation chain and take the reciprocal as the density value:

  \[
  \text{Density} = \frac{1}{\sum_{i=1}^{n} LOC_i}
  \]

  where $LOC_i$ denotes the number of lines of code in the $i$-th method of the invocation chain. A larger invocation chain corresponds to a smaller Density value, indicating that the target API occupies a smaller proportion of the overall implementation logic, thereby increasing the difficulty of recommendation.
\end{itemize}

These metrics enable quantification of both the invocation relationship between a query method and its target APIs and the relative importance of APIs within the implementation logic, providing a foundation for subsequent performance analysis of API recommendation methods.

The final \textbf{Q-ARE} dataset comprises 115 open-source projects, 5,698 query samples, and 1,822 distinct third-party APIs, with a maximum call depth of 7. The dataset systematically captures both direct and indirect API usage in real software projects, providing a new benchmark for evaluating API recommendation methods on tasks involving complex call structures and semantic reasoning.

\section{Experimental Setup and Results}
\subsection{Experimental Objectives}

The experiments in this work aim to leverage the \textbf{Q-ARE} dataset to systematically evaluate the performance of existing query-based API recommendation methods as well as general large language models (LLMs) under complex semantic invocation structures. The experimental design primarily focuses on:

\begin{itemize}
  \item Assessing the capabilities of baseline methods on the \textbf{Q-ARE} dataset.
  \item Systematically analyzing the relationships between the two metrics, \textbf{API Call Depth} and \textbf{API Call Density}, and semantic relevance:
  \begin{itemize}
    \item Investigating the impact of \textbf{API Call Depth} on recommendation performance.
    \item Investigating the impact of \textbf{API Call Density} on recommendation performance.
    \item Studying the relationship between \textbf{API Call Depth}, \textbf{API Call Density}, and semantic relevance.
  \end{itemize}
\end{itemize}

\subsection{Experimental Subjects and Methods}

To comprehensively evaluate performance on the \textbf{Q-ARE} dataset, we select three representative types of query-based API recommendation approaches as baselines, including retrieval-based methods, learning-based methods, and general LLM-based methods.

\textbf{Retrieval-based method.}  
Retrieval-based approaches recommend APIs by measuring the similarity between queries and API documentation or code examples. We adopt \textbf{CLEAR}~\cite{wei2022clear} as the baseline. CLEAR uses RoBERTa to generate sentence embeddings and employs a two-stage framework consisting of retrieval followed by BERT-based binary re-ranking to recommend APIs. We choose CLEAR because it is a representative retrieval-based method in existing studies and provides a reproducible implementation, ensuring reliable and fair evaluation.

\textbf{Learning-based method.}  
Learning-based approaches encode method call sequences or call graphs using sequence models or graph neural networks to predict target APIs. We adopt \textbf{DeepAPI}~\cite{gu2016deep} as the baseline. DeepAPI employs an encoder--decoder architecture with an attention mechanism to generate accurate API call sequences. We select DeepAPI as it is widely recognized in the literature and has a reproducible implementation, which guarantees the consistency and fairness of our experimental evaluation.

\textbf{Large Language Model.}  
In the LLM-based approach, method comments are used as natural language queries, and the model selects relevant APIs from a given API list. We adopt \textbf{Gemini-3-Pro-Preview} as the baseline model, which ranks among the top models on the OpenCompass LLM Leaderboard (January 2026)~\cite{OpenCompass2024}.

\subsection{Experimental Setup}

\subsubsection{Research Questions (RQs)}
Based on the experimental objectives of this work and the characteristics of the \textbf{Q-ARE} dataset, we define the following core research questions to systematically evaluate the performance of existing query-based API recommendation methods and general large language models (LLMs). These RQs cover both horizontal comparisons among methods and vertical analyses of metrics, with sub-questions focusing on the relationship between \textbf{API Call Depth}, \textbf{API Call Density}, and semantic relevance, providing clear guidance for experiment design and result analysis:

\begin{itemize}
    \item \textbf{RQ1}: How do different baseline methods perform on the \textbf{Q-ARE} dataset?
    \item \textbf{RQ2}: How do the two metrics, \textbf{API Call Depth} and \textbf{API Call Density}, affect API recommendation results, and what is their relationship with semantics?
\end{itemize}

RQ2 focuses on the two key metrics, \textbf{API Call Depth} and \textbf{API Call Density}, to systematically investigate their impact on recommendation performance and their semantic correlations. This is further decomposed into three sub-questions:

\begin{itemize}
    \item \textbf{RQ2.1}: How does \textbf{API Call Depth} influence API recommendation results?
    \item \textbf{RQ2.2}: How does \textbf{API Call Density} influence API recommendation results?
    \item \textbf{RQ2.3}: What is the relationship between \textbf{API Call Depth}, \textbf{API Call Density}, and semantic relevance?
\end{itemize}

\subsubsection{Evaluation Metrics}
To systematically assess the performance of query-based API recommendation methods on the \textbf{Q-ARE} dataset, we adopt multiple evaluation metrics to comprehensively measure the quality of recommendations. Following the approach of ~\cite{wei2022clear}, we employ three widely-used metrics: Precision, Recall, and F1-score. Additionally, we introduce the API Exact Match (AEM) metric proposed in ~\cite{lin2025lightweight}, which evaluates whether the recommended results exactly match the target APIs.

\subsection{Experimental Results and Analysis}

\subsubsection{RQ1: Performance of Different Baseline Methods on the Q-ARE Dataset}

\noindent First, we compute and report the Precision, Recall, and API Exact Match for the three selected baselines on the \textbf{Q-ARE} dataset. The results are presented in Table~\ref{tab:baseline_performance}.

\begin{table}[ht]
\centering
\caption{Performance of baseline methods on the Q-ARE dataset}
\label{tab:baseline_performance}
\begin{tabular}{lccc}
\toprule
\textbf{Method} & \textbf{Precision} & \textbf{Recall} & \textbf{API Exact Match} \\
\midrule
DeepAPI & 0.00009 & 0.00018 & 0.01738 \\
CLEAR & 0.02229 & 0.01331 & 0 \\
Gemini-3-Pro-Preview & 0.21000 & 0.20140 & 0.10760 \\
\bottomrule
\end{tabular}
\end{table}

\sloppy
As shown in Table~\ref{tab:baseline_performance}, when confronted with complex scenarios in which methods and APIs in real software projects are connected through multi-level invocation chains, the overall performance of existing API recommendation methods remains limited. Regarding the API Exact Match metric, \textbf{Gemini-3-Pro-Preview} achieves the best result, yet its match rate is only around 11\%. This indicates that even state-of-the-art general large language models still face significant challenges in accurately predicting the complete set of target APIs under complex invocation structures.

The primary reason for this phenomenon is that, in real software development projects, target APIs are likely not invoked directly (single-level) by the query method, but rather through multiple intermediate method calls. As the call depth increases, recommendation models struggle to capture the true usage intent of the APIs. Moreover, a single method implementation typically involves the collaborative usage of multiple APIs, further increasing the difficulty of predicting the complete set of APIs.

Overall, these results indicate that current API recommendation methods still have substantial limitations in handling complex invocation structures, highlighting the necessity of incorporating invocation structure information into API recommendation tasks.

\subsubsection{RQ2: How do \textbf{API Call Depth} and \textbf{Call Density} affect recommendation results, and what is their relationship with semantics?}

This research question focuses on the two key metrics, \textbf{API Call Depth} and \textbf{API Call Density}, systematically investigating their impact on API recommendation performance and their relationship with semantic understanding. Specifically, it is further decomposed into the following three sub-questions:

\textbf{RQ2.1: How does \textbf{API Call Depth} affect API recommendation results?}

\begin{figure}
\centering
\includegraphics[width=0.7\linewidth]{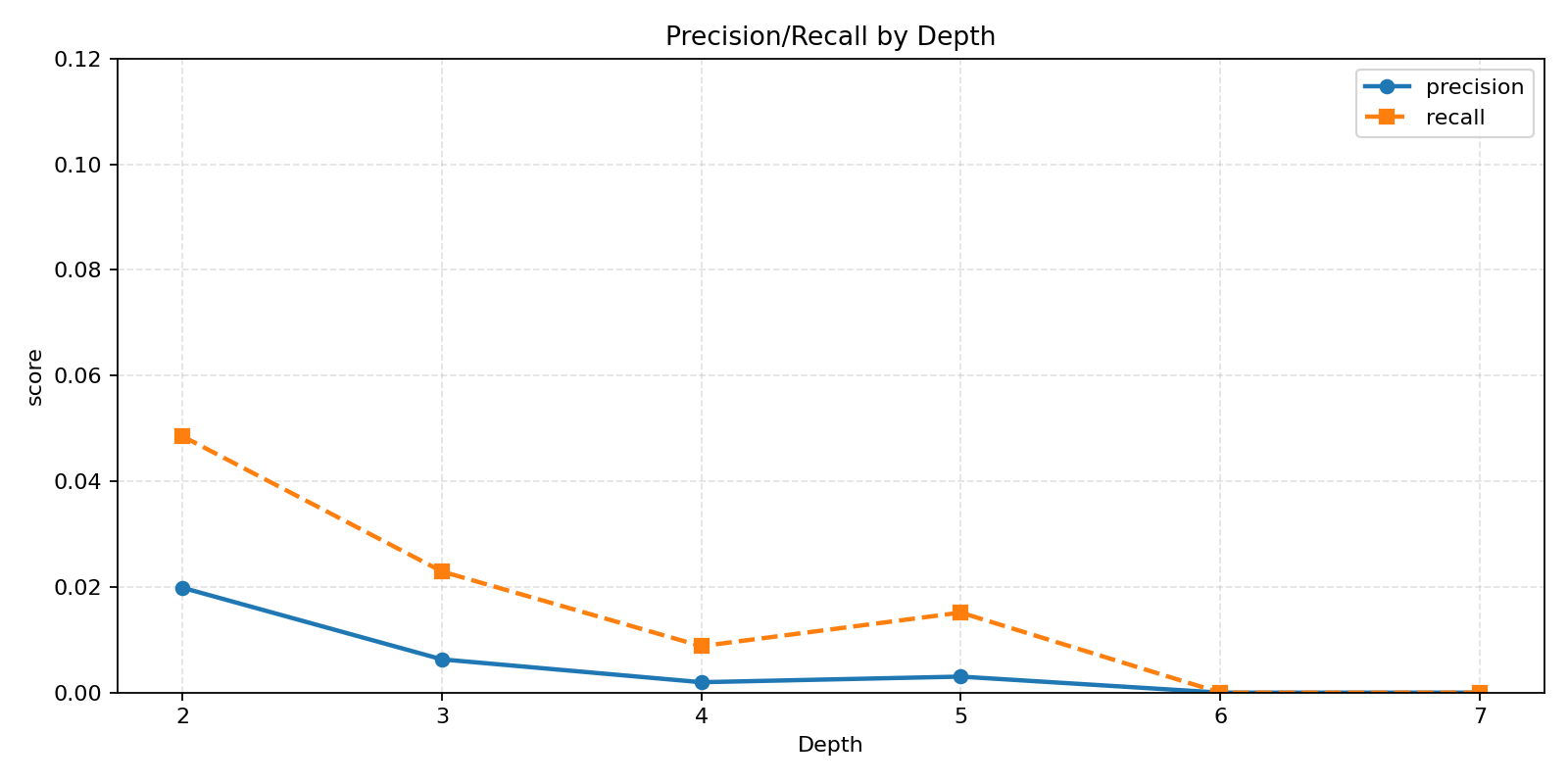}
\\[1ex]
\textbf{(a) DeepAPI}

\includegraphics[width=0.7\linewidth]{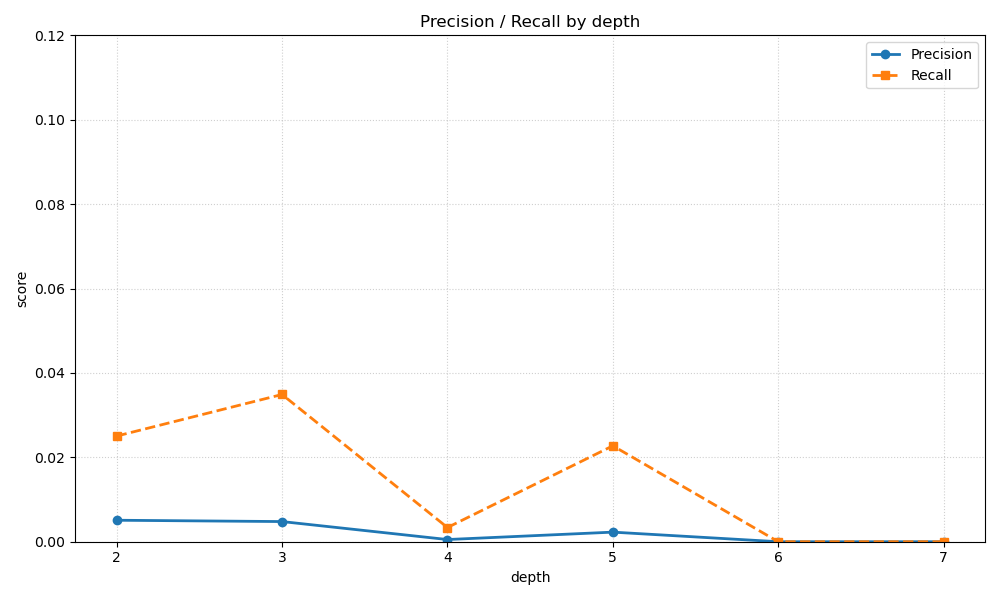}
\\[1ex]
\textbf{(b) CLEAR}

\includegraphics[width=0.7\linewidth]{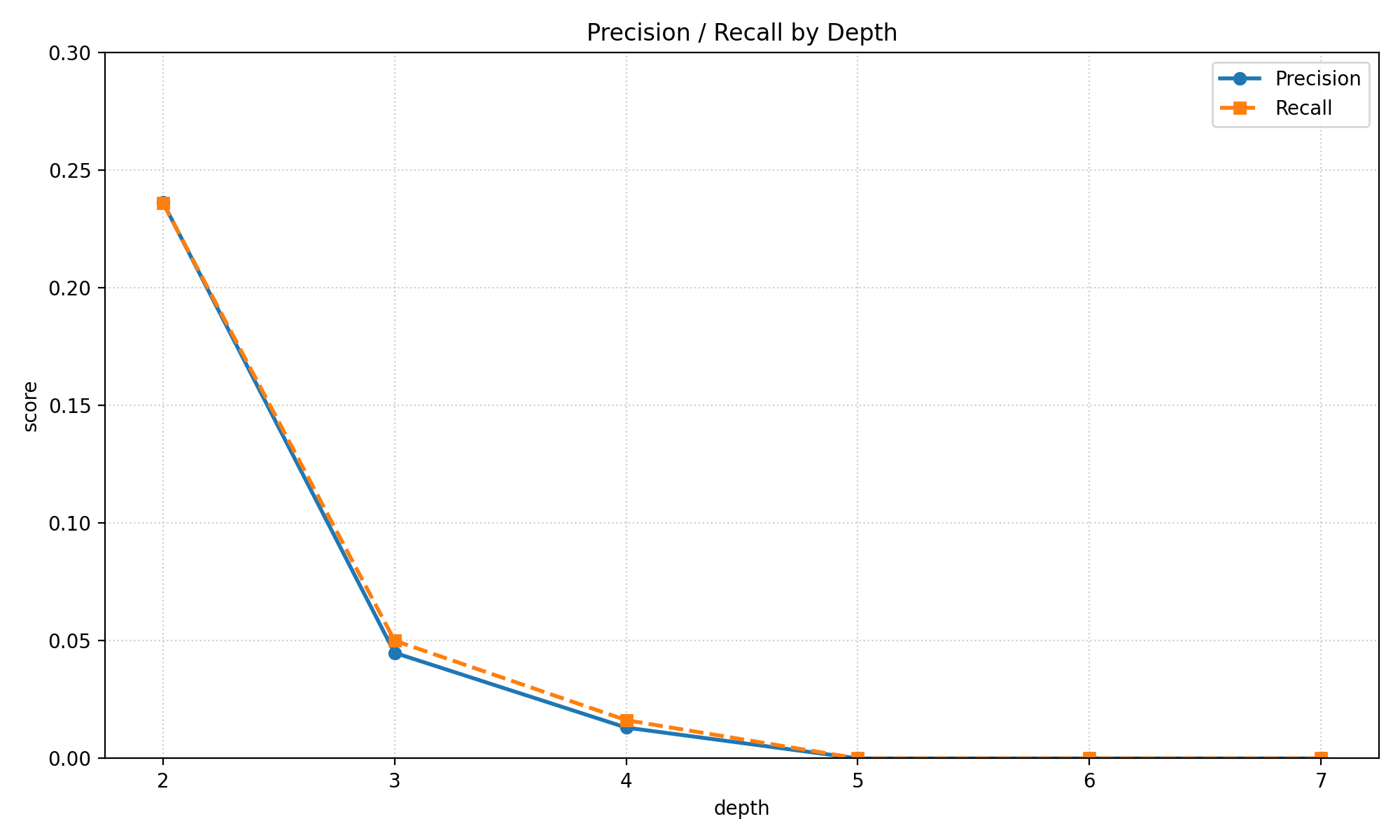}
\\[1ex]
\textbf{(c) Gemini-3-Pro-Preview}

\caption{Performance comparison of baseline methods under different API Call Depths}
\label{fig:depth_comparison}
\end{figure}

For this sub-question, we analyze the precision, recall, and F1 scores of the three baseline methods under different \textbf{API Call Depths}. The corresponding line charts are presented in Figure~\ref{fig:depth_comparison}.

The experimental results indicate that all three methods achieve their best performance when the \textbf{API Call Depth} is 2. As the invocation chain deepens, both precision and recall scores decrease significantly. Among them, \textbf{CLEAR} exhibits some fluctuations at depths 3 and 5, which may be caused by the uneven distribution of samples across different depths. \textbf{DEEPAPI} shows a relatively smooth performance decline, while \textbf{Gemini} performs well for shallow calls but remains ineffective for deep calls. Overall, existing methods still face substantial challenges in handling deep API calls, highlighting the need for more effective API recommendation approaches to improve prediction capability in scenarios with multi-level invocation chains.

\textbf{RQ2.2: How does \textbf{API Call Density} affect API recommendation performance?}  

\begin{figure}
\centering
\includegraphics[width=0.7\linewidth]{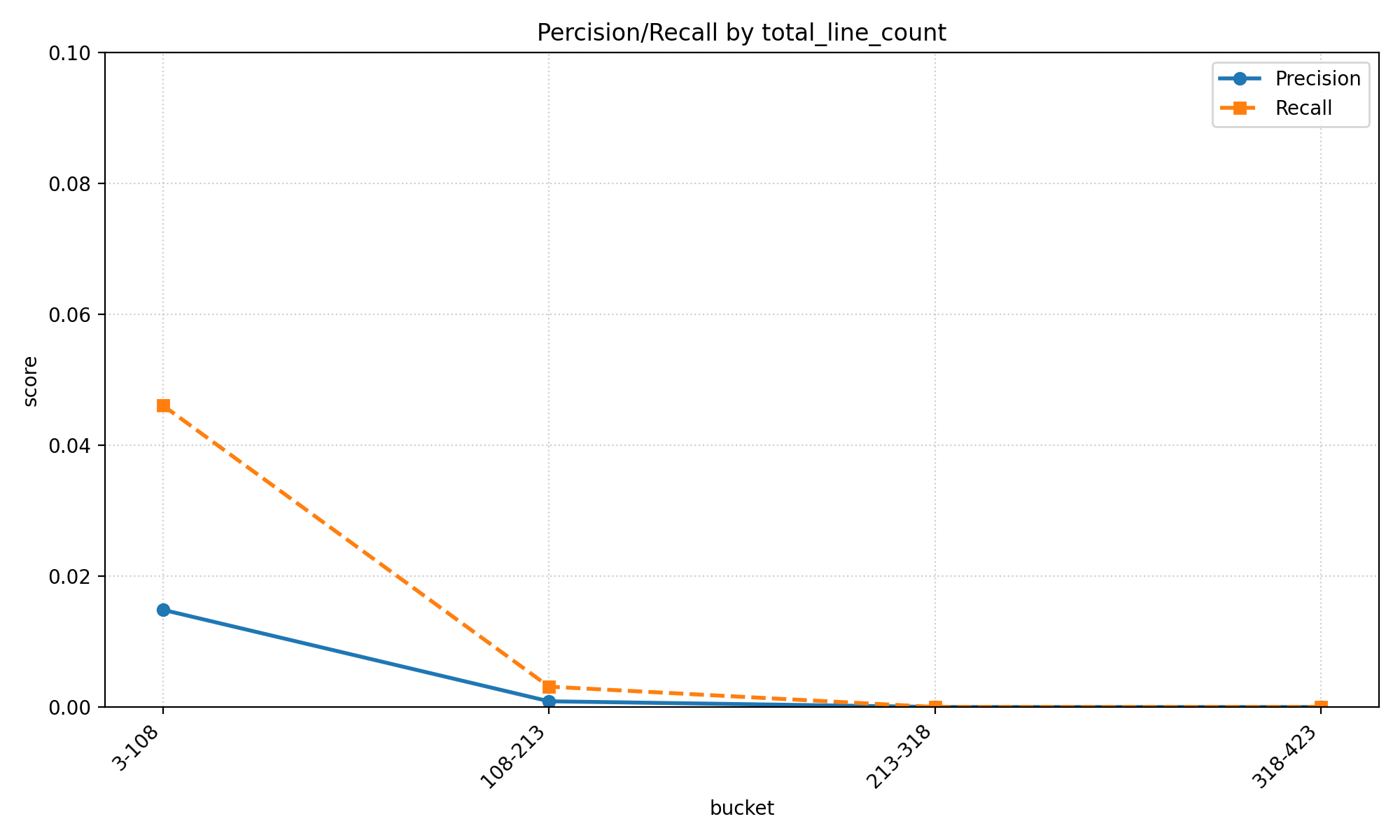}
\\[1ex]
\textbf{(a) DeepAPI}

\includegraphics[width=0.7\linewidth]{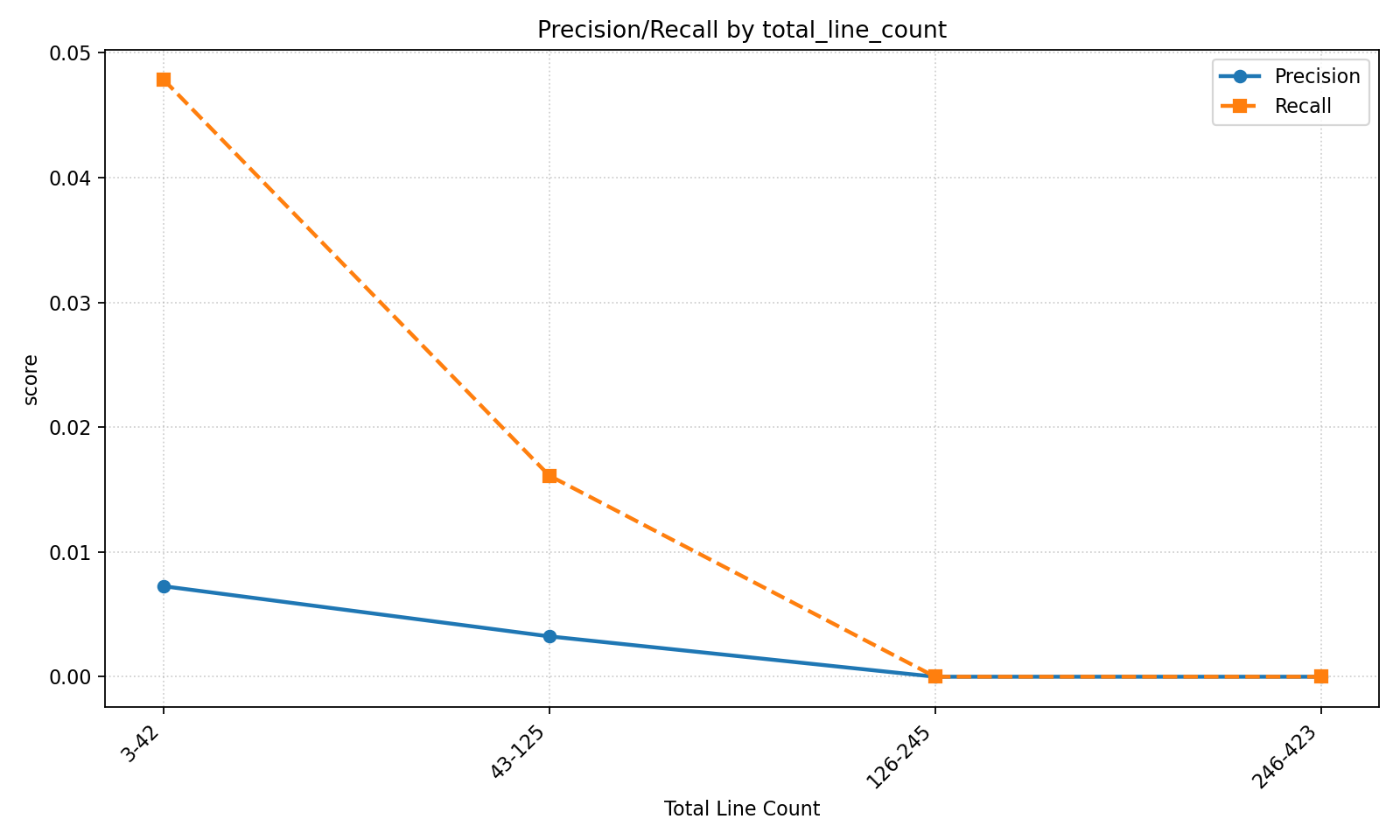}
\\[1ex]
\textbf{(b) CLEAR}

\includegraphics[width=0.7\linewidth]{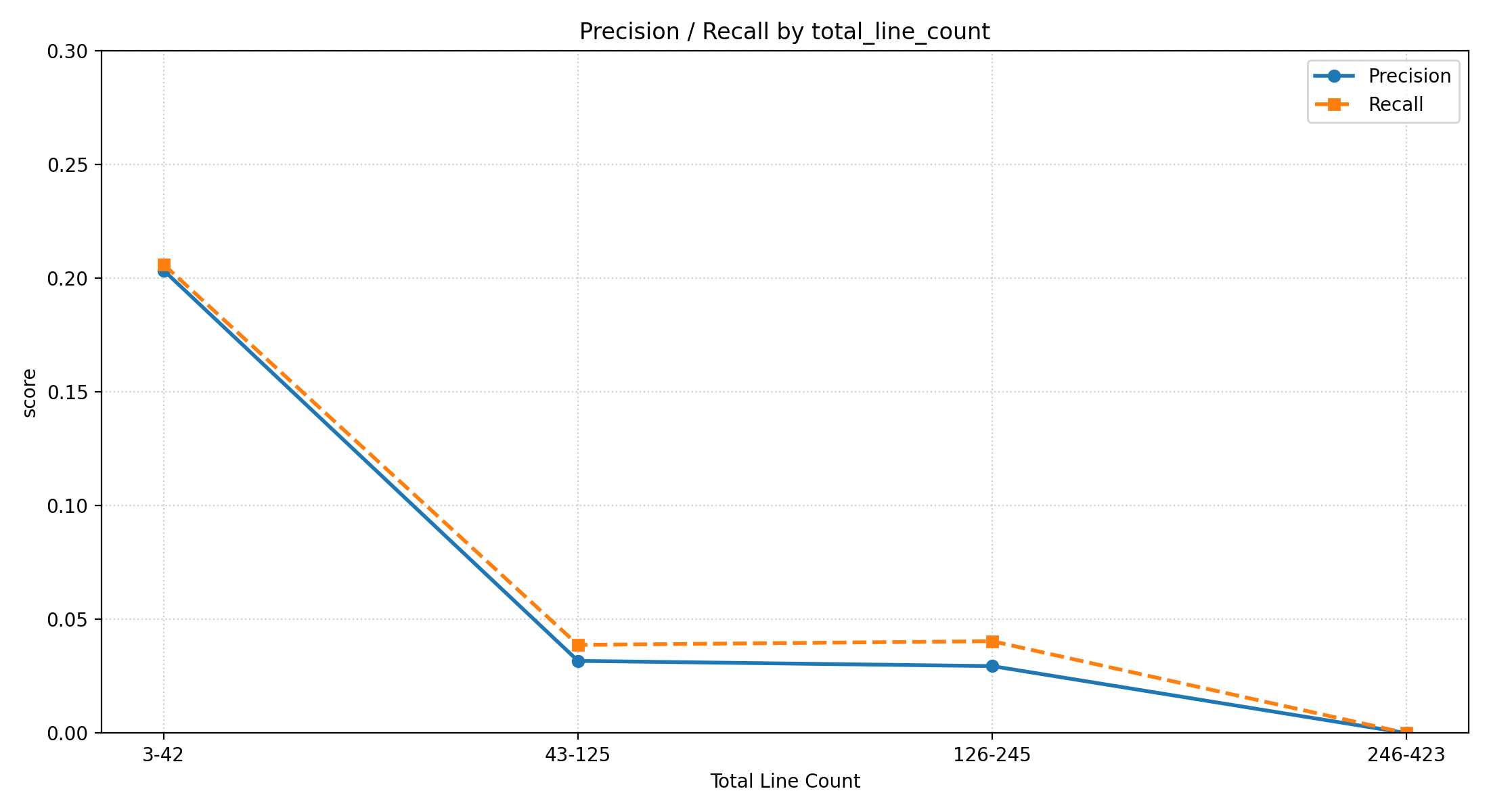}
\\[1ex]
\textbf{(c) Gemini-3-Pro-Preview}

\caption{Performance comparison of baseline methods under different method total line counts (Density)}
\label{fig:density_comparison}
\end{figure}

\textbf{API Call Density} is defined as the reciprocal of the total lines of code of a method. We further analyze the precision and recall scores of the three baseline methods under different total lines of code, and plot the corresponding line charts, as shown in Figure~\ref{fig:density_comparison}.

The experimental results indicate that all methods exhibit significant \emph{length sensitivity}. For shorter code segments (less than 125 lines), the models retain a certain level of recognition capability; however, as the length of method implementations increases, performance drops substantially, and for code segments exceeding 245 lines, the models are nearly ineffective.

Specifically, \textbf{DeepAPI} achieves relatively high precision on short code segments but suffers from low recall, indicating conservative predictions that may miss true target APIs. \textbf{CLEAR} demonstrates slightly higher recall on short code segments but lower precision, resulting in more false positives. \textbf{Gemini-3-Pro-Preview}, on the other hand, exhibits the highest precision on short code segments (up to 0.2), outperforming the other two methods and highlighting the potential of large language models in local semantic understanding and contextual modeling.

Nevertheless, all methods perform poorly on long code segments, indicating that existing techniques still face significant challenges when handling complex and lengthy invocation structures. Therefore, future research should focus on improving the performance of API recommendation methods for long code segments to better handle complex invocation chains in real-world software projects.

\textbf{RQ2.3: Relationship between API Call Depth, Density, and Semantic Relevance}

\textbf{Analysis of API Call Depth and Semantic Relevance:}  
For each query sample, we record the \textbf{Depth} of its corresponding target APIs and group the samples by Depth. Precision and Recall are then calculated for each Depth group to observe the impact of invocation distance on recommendation performance. Additionally, to validate this RQ, we compute the semantic similarity between the query method comments and the target API documentation for each Depth group, and visualize the trend using line charts.

\textbf{Analysis of API Call Density and Semantic Relevance:}  
For each query sample, we record the \textbf{Density} of its corresponding target APIs. Density values are divided into six groups ranging from 0 to 0.333333. Precision and Recall are calculated for each \textbf{Depth}group within each Density interval to analyze the impact of API density on recommendation performance. Similarly, to validate this RQ, we compute the semantic similarity between query method comments and the corresponding target API documentation within each Density group, and visualize the results using bar charts.

To answer this research question, we employ Sentence-Transformers to compute the semantic similarity between each query method comment and its corresponding third-party API documentation under different Depth values and Density intervals. The results are visualized using line and bar charts to investigate the correlation between Depth, Density, and semantic similarity, thereby validating the effectiveness of these two metrics in capturing the semantic relevance between queries and target APIs.

\begin{figure}[t]
\centering
\includegraphics[width=0.7\linewidth]{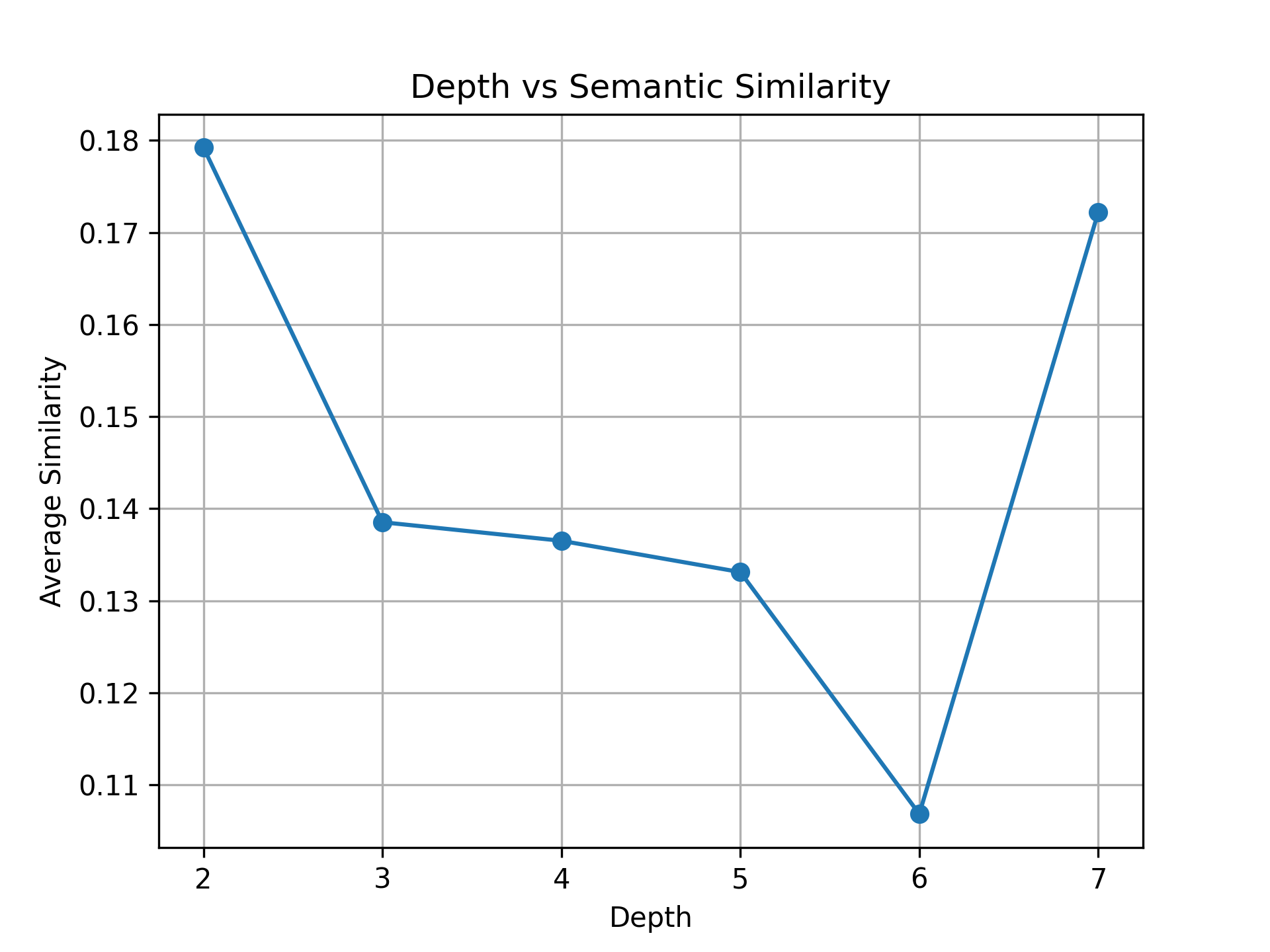}
\caption{Semantic similarity across different \textbf{Depth} groups}
\label{fig:depth_semantic}
\end{figure}

Figure~\ref{fig:depth_semantic} illustrates the trend of semantic similarity between query comments and target API documentation with respect to \textbf{API Call Depth}. The x-axis represents Depth values (ranging from 2 to 7, where Depth=2 indicates direct calls, and larger values indicate more distant invocations), and the y-axis represents the average semantic similarity at each depth.

From the figure, it can be observed that \textbf{Depth} exhibits a clear negative correlation with semantic similarity. As \textbf{Depth} increases (i.e., the invocation distance between the query method and the target API becomes longer), the average semantic similarity generally decreases. When \textbf{Depth} = 2 (direct invocation), the average semantic similarity reaches its maximum value (approximately 0.18), indicating the strongest semantic association between the query method and directly invoked APIs. When \textbf{Depth} ≥ 3, semantic similarity significantly drops, suggesting that the semantic connection between the query and target APIs weakens as the invocation chain length increases. 

It is worth noting that at \textbf{Depth} = 7, there is a slight increase in semantic similarity. Further analysis reveals that the number of samples in this \textbf{Depth} interval is only 11, which may lead to statistical fluctuations. Moreover, several of these samples have Density values higher than the average Density for \textbf{Depth} 5 and 6. Overall, these results indicate that \textbf{Depth} effectively reflects the semantic distance between query methods and target APIs, providing preliminary evidence for its validity as a metric for semantic relevance.

\begin{figure}[t]
\centering
\includegraphics[width=0.85\linewidth]{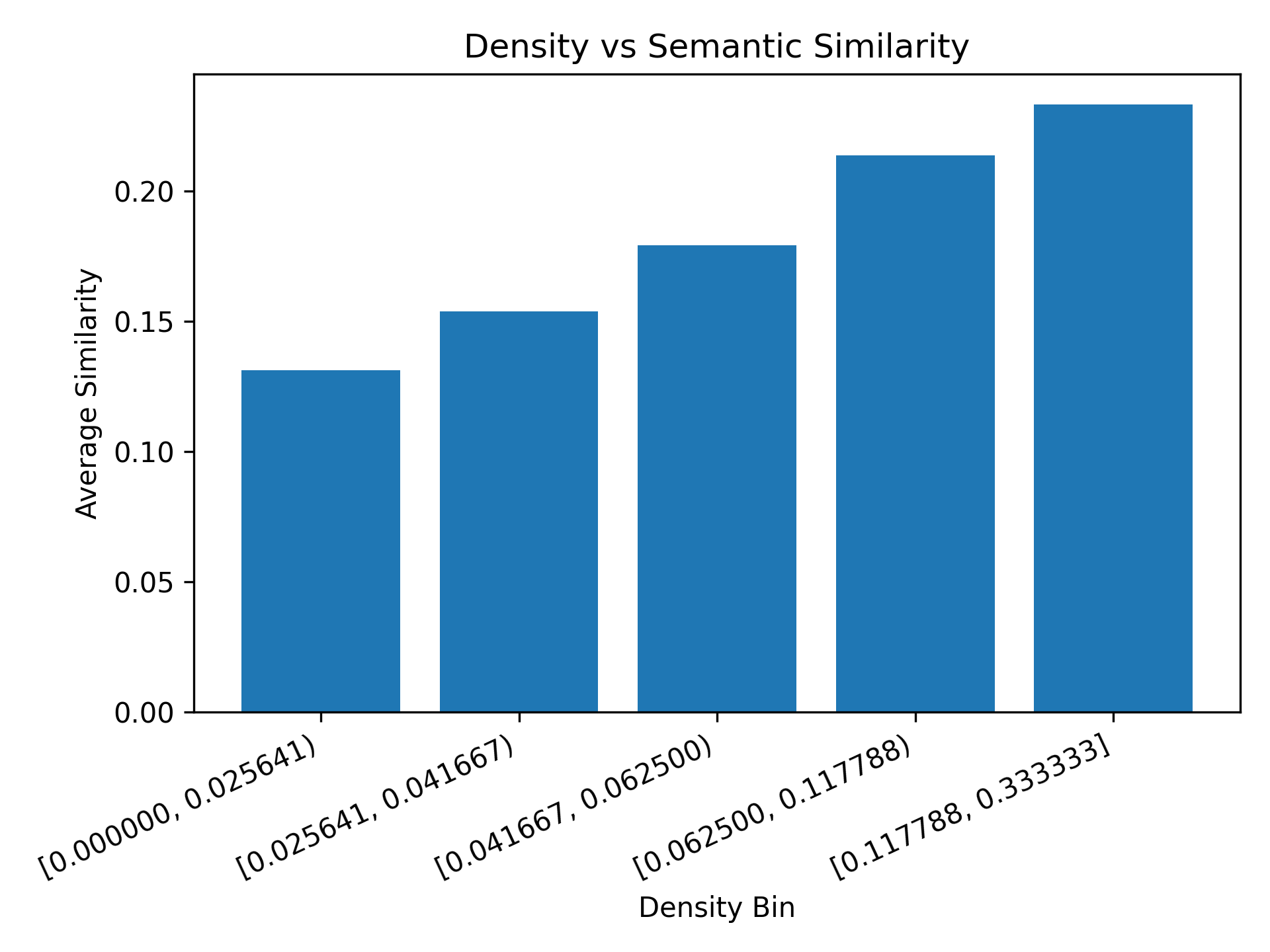}
\caption{Semantic similarity across different Density groups}
\label{fig:density_semantic}
\end{figure}

Figure~\ref{fig:density_semantic}illustrates the comparison of average semantic similarity between query comments and target API documentation under different Density groups. The x-axis represents the Density intervals, while the y-axis shows the average semantic similarity for each group.

As shown in Figure~\ref{fig:density_semantic}, semantic similarity exhibits a clear positive correlation with \textbf{Density}. As Density increases (i.e., the semantic proportion of the target API within the invocation chain code rises), the average semantic similarity between the query method and the target API also increases. Specifically, the high-Density group achieves an average semantic similarity of approximately 0.23, representing an increase of about 0.10 compared to the low-Density group. 

These results indicate that when the target API occupies a higher semantic proportion in the method implementation, the query method's semantic information more effectively reflects the API usage intent, resulting in stronger semantic alignment. Therefore, Density effectively captures the semantic importance of the target API within the invocation chain and further validates its close relationship with query–API semantic relevance.

Based on the above statistical analysis and visualizations, RQ2.4 can be addressed as follows:  
\begin{itemize}
    \item \textbf{Depth} is generally negatively correlated with semantic similarity: the greater the invocation \textbf{Depth}, the weaker the semantic association between the query method and the target API, and the lower the semantic proportion of the target API in the method implementation.
    \item \textbf{Density} is positively correlated with semantic similarity: the higher the invocation density, the stronger the semantic association between the query method and the target API, and the higher the semantic proportion of the target API in the method implementation.
\end{itemize}

Both \textbf{Depth} and Density can to some extent reflect the semantic relevance between the query and the target API. Hence, these two metrics serve as important factors for analyzing the difficulty of API recommendation under complex semantic scenarios, providing a new analytical dimension for evaluating the performance of API recommendation methods.

\subsection{Experimental Conclusions}

This chapter systematically evaluates the performance of existing query-based API recommendation methods and general LLMs on the QARE dataset, verifying the research questions through statistical analysis and visualization. The main conclusions are:

\begin{enumerate}
    \item \textbf{Performance of traditional methods vs. LLMs:} LLMs perform better in single-layer invocations, while traditional methods better capture invocation chain structures. However, all methods struggle with multi-layer invocations and low-density scenarios, indicating room for improvement in semantic understanding and structural awareness.
    
    \item \textbf{Impact of Depth and Density:} As \textbf{Depth} increases, precision drops sharply, showing a strong negative effect on recommendation performance. Density generally has a positive effect, while deeper structures remain challenging.
    
    \item \textbf{API call metrics and semantic relevance:} \textbf{Depth} negatively correlates with semantic similarity between query and target API, while Density shows a positive correlation. Both metrics effectively characterize semantic relevance, offering a new dimension for evaluating API recommendation performance.
\end{enumerate}

\section{Discussion and Future Work}

Experimental results show that existing query-based API recommendation methods and general-purpose LLMs achieve only around 11\% AEM in complex software scenarios, indicating overall limited performance. As \textbf{API Call Depth} increases and \textbf{API Call Density} decreases—reducing the semantic proportion of APIs—all baseline methods’ performance drops significantly. This highlights the limitations of current approaches in capturing multi-layered invocation relationships and complex semantic dependencies. Moreover, the strong impact of \textbf{Call Density} shows that existing methods struggle to capture semantic relevance in sparse invocation scenarios, making it difficult to exploit latent semantic relationships in indirect call chains.

The \textbf{Q-ARE} dataset introduces two evaluation dimensions—\textit{Depth} and \textit{Density}—addressing the limitations of traditional datasets that consider only direct query-to-API matches. This provides a finer-grained, realistic benchmark for evaluating API recommendation methods under complex scenarios. Experimental results validate the dataset and metrics, while revealing key challenges in handling multi-layered invocation chains: balancing semantic comprehension (\textbf{Depth}) with structural complexity. These insights indicate directions for future research.

Future work can focus on:
\begin{itemize}
\item \textbf{Enhancing multi-layer call reasoning:} Develop more efficient reasoning mechanisms to better capture multi-layered chains and semantic dependencies.
\item \textbf{Expanding dataset scale and coverage:} Build larger, multi-language, multi-framework datasets to overcome ecological limitations and improve generalizability.
\item \textbf{Integrating multimodal information and improving annotations:} Combine source code, documentation, and test cases, while improving LLM-generated annotations to reduce semantic noise.
\item \textbf{Designing multi-metric recommendation strategies:} Use \textit{Depth} and \textit{Density} to optimize ranking, improving accuracy in indirect or sparse invocation scenarios.
\end{itemize}

These improvements are expected to enhance API recommendation accuracy and semantic understanding in complex projects, reducing developer effort and supporting more efficient software development.

\section{Threats to Validity}

To ensure the reliability of our conclusions, this study systematically analyzes potential threats to validity in dataset construction, experimental design, and evaluation, and adopts measures to mitigate them.  

Selection bias may arise because the \textbf{Q-ARE} dataset only includes Java projects with at least 20 stars, test cases, and Spring Framework development, which could limit diversity in project type, scale, and coding style. To mitigate this, we included projects from multiple domains with diverse third-party library usage and multi-layered APIs covering both direct and indirect calls, better reflecting real-world software and reducing sample bias.  

The accuracy of call chain analysis is another potential threat. Indirect API calls were identified via recursive static analysis of method call chains, which may miss or misparse some calls, affecting evaluation accuracy. We mitigated this risk by filtering the parsed call chains and excluding APIs that lacked source code or annotations.  

Annotation quality and semantic noise also affect evaluation. Although the \textbf{Q-ARE} dataset combines method comments, official Javadoc, and LLM-generated descriptions, some comments may be low-quality and LLM summaries may be inaccurate, introducing semantic noise that could misestimate recommendation performance. To address this, we use only the first two lines of method comments, combine official documentation with source context, and filter low-quality or missing annotations.  

Finally, limitations in evaluation metrics and experimental settings pose a threat. Our experiments use Precision, Recall, and AEM, focusing on \textbf{call depth} and \textbf{density}, but factors such as API usage frequency, context, and runtime behavior are not considered, which may underestimate some methods. Future work can incorporate dynamic analysis and multimodal information for a more comprehensive evaluation.  

In summary, these measures collectively minimize threats to validity, providing a reliable basis for evaluating API recommendation methods using \textbf{Q-ARE} and guiding future related research.

\section{Conclusion}

This work focuses on query-based API recommendation and introduces a new evaluation dataset, \textbf{Q-ARE}. The dataset extracts methods and their associated third-party APIs from real open-source Java projects, and recursively parses method call chains to construct target API sets. Additionally, we introduce the concepts of \textbf{API Call Depth} and \textbf{API Call Density}, quantifying:

\begin{itemize}
    \item the distance between query methods and target APIs, and
    \item the compactness of API semantic contributions within call chains,
\end{itemize}

thereby providing new evaluation dimensions for analyzing recommendation performance in complex call scenarios. To support semantic understanding assessment, \textbf{Q-ARE} includes rich API annotations, including official Javadoc as well as summaries generated from source code analysis and large language models.

Experimental results based on the \textbf{Q-ARE} dataset indicate:

\begin{enumerate}
    \item As \textbf{API Call Depth} increases and call chains become sparser, the performance of existing query-based methods and general-purpose LLMs significantly decreases, revealing the limitations of current approaches in cross-method call reasoning and complex semantic understanding.
    \item The \textbf{Q-ARE} dataset and evaluation metrics effectively support performance analysis across different types of methods, providing a practical platform for standardized API recommendation evaluation and serving as a reference for method improvement.
\end{enumerate}

From a practical perspective, the dataset is not only suitable for academic evaluation of API recommendation and code generation models but also provides direct support for industrial development tools. In practice, Q-ARE can be used to train and evaluate IDE-based intelligent programming assistants, such as API recommendation and code completion models, improving their accuracy in real project contexts and, consequently, developer productivity. Furthermore, the extracted method and API invocation relationships can support software engineering tasks such as code comprehension, dependency analysis, and project-level API usage pattern mining. Therefore, Q-ARE holds value for both academic research and industrial applications.

Overall, this study provides new data resources, evaluation dimensions, and experimental infrastructure for the field of API recommendation. It facilitates systematic evaluation of current methods’ limitations and lays the foundation for future research, including enhancing cross-method reasoning capabilities of LLMs and designing more robust semantic-aware recommendation mechanisms. Future work may explore efficient multi-level call chain modeling methods and leverage LLMs’ contextual understanding capabilities to continuously improve the accuracy and practicality of API recommendation.

\bibliographystyle{ACM-Reference-Format}
\bibliography{reference}

\end{document}